\documentclass[[journal=nalefd,manuscript=letter]{achemso}

\usepackage{graphicx}
\usepackage{amsmath}
\usepackage{dcolumn}   
\usepackage{bm}        
\usepackage{amssymb}   
\usepackage{graphicx,epsfig}
\usepackage{color}
\usepackage[usenames,dvipsnames]{xcolor}
\usepackage[ocgcolorlinks,colorlinks=true,linkcolor=blue,citecolor=red]{hyperref}
\usepackage{amsmath,amssymb, amsthm, amsfonts}
\usepackage{xspace}
\usepackage{xcolor}
\usepackage{verbatim}
\usepackage{mathtools}

\title{Electrostatic field-driven supercurrent suppression in ionic-gated metallic superconducting nanotransistors}

\author{Federico Paolucci}
\email{federico.paolucci@nano.cnr.it}
\affiliation{NEST, Istituto Nanoscienze-CNR and Scuola Normale Superiore, Piazza San Silvestro 12, I-56127 Pisa, Italy}

\author{Francesco Cris\'a}
\affiliation{Department  of  Physics  “E. Fermi”,  Universit\'a  di  Pisa,  Largo  Pontecorvo  3, I-56127 Pisa, Italy}

\author{Giorgio De Simoni}
\affiliation{NEST, Istituto Nanoscienze-CNR and Scuola Normale Superiore, Piazza San Silvestro 12, I-56127 Pisa, Italy}

\author{Lennart Bours}
\affiliation{NEST, Istituto Nanoscienze-CNR and Scuola Normale Superiore, Piazza San Silvestro 12, I-56127 Pisa, Italy}

\author{Claudio Puglia}
\affiliation{NEST, Istituto Nanoscienze-CNR and Scuola Normale Superiore, Piazza San Silvestro 12, I-56127 Pisa, Italy}

\author{Elia Strambini}
\affiliation{NEST, Istituto Nanoscienze-CNR and Scuola Normale Superiore, Piazza San Silvestro 12, I-56127 Pisa, Italy}

\author{Stefano Roddaro}
\affiliation{Department  of  Physics  “E. Fermi”,  Universit\'a  di  Pisa,  Largo  Pontecorvo  3, I-56127 Pisa, Italy}

\author{Francesco Giazotto}
\email{francesco.giazotto@sns.it}
\affiliation{NEST, Istituto Nanoscienze-CNR and Scuola Normale Superiore, Piazza San Silvestro 12, I-56127 Pisa, Italy}

\abbreviations{IR,NMR,UV}

\begin{document}

\begin{tocentry}
\center
\includegraphics  [height=4.45cm]{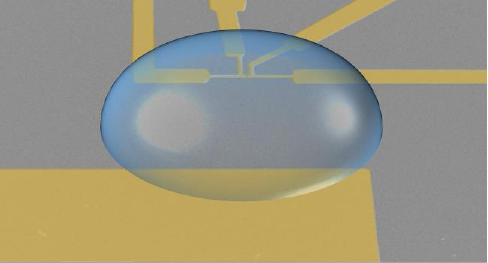}

\end{tocentry}

\begin{abstract}
Recent experiments have shown the possibility to tune the transport properties of metallic nanosized superconductors through a gate voltage. These results renewed the longstanding debate on the interaction between electrostatic fields and superconductivity. Indeed, different works suggested competing mechanisms as the cause of the effect: unconventional electric field-effect or quasiparticle injection. 
Here, we provide the conclusive evidence of electrostatic field-driven control of the supercurrent in metallic nanosized superconductors, by realizing ionic-gated superconducting field-effect nanotransistors (ISFETs) where electron injection is impossible. Our Nb ISFETs show giant suppression of the superconducting critical current up to $\sim45\%$.
Moreover, the bipolar supercurrent suppression observed in different ISFETs, together with invariant critical temperature and normal-state resistance, exclude also conventional charge accumulation/depletion.
Therefore, the microscopic explanation of this effect calls upon a novel theory able to describe the non-trivial interaction of static electric fields with conventional superconductivity.
\end{abstract}

Since the discovery of superconductivity, the impact of electrostatic fields on metallic superconductors is an open issue. Seminal experiments showed modulations of the superconducting critical temperature ($T_C$) and normal-state resistance ($R_N$) in tin and indium films \cite{Glover1960,Stadler1965} by conventional charge accumulation/depletion. Recently, gate voltage ($V_g$) induced \textit{bipolar} full supercurrent suppression was demonstrated via conventional solid gating in several metallic superconducting nanostructures, such as wires \cite{DeSimoni2018,Orus2021}, constriction Josephson junctions \cite{Paolucci2018}, and proximitized normal metals \cite{DeSimoni2019}. Similar results were obtained also on suspended \cite{Rocci2020} and epitaxial \cite{Elalaily2021} superconducting metal layers grown on semiconductor nanowires. 
Differently from charge accumulation/depletion, this gating effect seems also to influence the phase of the Cooper pairs condensate \cite{Paolucci2019_2,Puglia2020,DeSimoni2021} without apparent impact on $T_C$ and $R_N$ \cite{Paolucci2019}, thereby suggesting a non-trivial interaction between electrostatic fields and superconductivity. 
Yet, subsequent works showed gate induced critical current ($I_C$) suppression in metallic superconductor nanostructures explained in terms of overheating due to quasiparticle injection into the transistor channel \cite{Ritter2021,Catto2021}, cold electron field-emission either from the gate electrode or the channel \cite{Golokolenov2021,Alegria2021} and non-equilibrium phonon-mediated interaction \cite{Ritter2021_2}. Differently, a recent experiment attributed the supercurrent suppression to current injection from the gate electrode inducing large energy fluctuations without any overheating \cite{Basset2021}. Despite quasiparticle injection can account for the critical current suppression, all these works are clearly unable to reconcile together all previous experimental evidences. Moreover, any mechanism that only destroys superconductivity is not able to account for the $\sim 30\%$ enhancement of the critical current observed in NbN nanobridges subject to conventional gating \cite{Rocci2021}.
Therefore, the microscopic origin of this phenomenon is still unknown and under a strong scientific debate. Beyond the profound relevance for fundamental science, electric field control of supercurrents in metallic superconductors would enable the implementation of a groundbreaking low-dissipation and energy-efficient superconducting electronics. 

\begin{figure*} [t!]
        \begin{center}
                \includegraphics{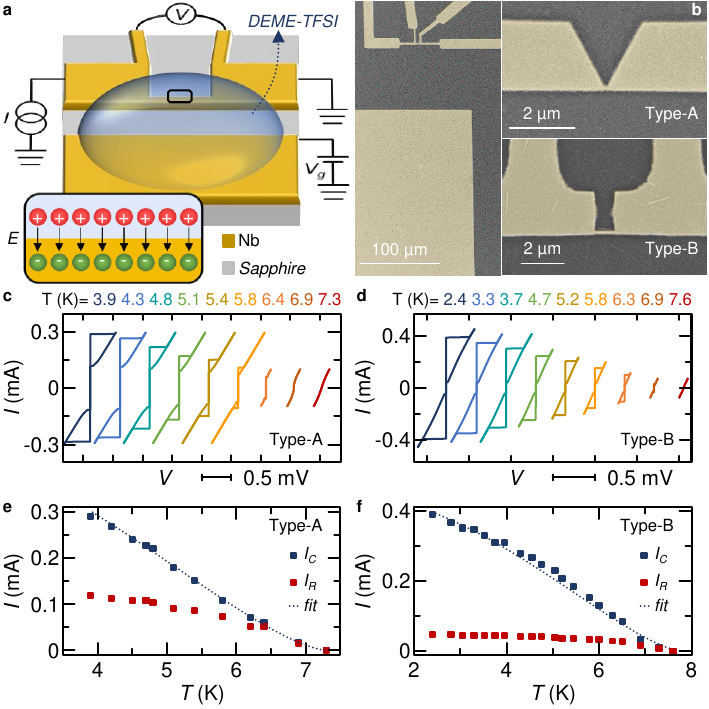}
        \end{center}
        \caption{\label{Fig1}\textbf{Transport properties of typical Nb ISFETs.} 
        (\textbf{a}) Schematic of a typical ISFET, where the superconducting active region is current biased, and the counter electrode polarizes the electrolyte (DEME-TFSI) droplet through the voltage $V_g$.  The accumulation of ions at the sample surface creates the EDL, and generates the electric field ($E$).
        (\textbf{b}) False-color SEM picture of typical ISFETs. (Top) Type-A device realized in the form of a Dayem bridge-like Josephson junction (the constriction length is $\sim 300$ nm while its width is $\sim 100$ nm). (Bottom) Type-B device consisting of a quasi one-dimensional superconducting nanowire (the wire length is $\sim 1\, \mu$m and its width is $\sim 100$ nm).
        (\textbf{c} and \textbf{d}) Current versus voltage ($IV$) characteristics for Type-A (c) and Type-B (d) ISFETs measured at different values of bath temperature. The curves are horizontally offset for clarity.
        (\textbf{e} and \textbf{f}) Temperature dependence of $I_C$ (blue) and $I_R$ (red) for Type-A (e) and Type-B (f) ISFETs. Dotted lines are the Bardeen theoretical curves for $I_C$. 
        }
\end{figure*}

Here, we provide the unequivocal demonstration of electrostatic field-induced giant $I_C$ suppression in metallic superconductors by exploiting ionic-gated Nb superconducting field-effect nanotransistors \cite{Nishino1989} (ISFETs, Fig. \ref{Fig1}a). By applying $V_g$ across an electrolyte (DEME-TFSI), ionic gating enables the generation of intense electrostatic fields ($E$) by forming the so-called electric double layer (EDL) at the sample surface (inset), in the absence of electron leakage currents. Indeed, electron transfer through an EDL is only possible by means of a redox reaction between the ions in the electrolyte and the surface atoms in the transistor channel \cite{Kirby2013,Calhoun1996,Boroda1997,Waegele2019}. This electrochemical interaction can only occur above the glass transition temperature of the electrolyte ($T_g$), since under this temperature the ions are immobile thus preventing electrolyte/electrode chemical interaction. For DEME-TFSI, the glass transition temperature is $T_g=182$ K \cite{McCann2015}, thus in our experiments no current can flow across the electrolyte/transistor channel interface. Furthermore, DEME-TFSI was employed in experiments very sensitive to current injection. Indeed, Coulomb blockade diamonds in InAs quantum dots \cite{Shibata2013} and quantized transport in SrTiO$_3$ superconducting quantum point contacts \cite{Mikheev2021} would be unattainable even in the presence of the quantized injection of a few single electrons from the electrolyte to the device.

Electrolyte gating is routinely used to tune the transport properties of micro-scale metals \cite{Daghero2012} and metallic superconductors \cite{Choi2014,Piatti2021} (i.e., $R_N$ and $T_C$) by charge accumulation/depletion, as it occurs in conventional semiconductor devices. Instead, we downsized to the nano-scale the channel of Nb ISFETs (Fig. \ref{Fig1}b) to study the impact of $E$ on the supercurrent flow, i.e., by measuring their $I_C$ versus $V_g$ characteristics. 
In particular, we realized different geometries by sputtering a niobium film on a sapphire substrate: Type-A devices are Josephson nanoconstrictions realized by a bottom-up approach, while Type-B samples are superconducting nanowires fabricated by a top-down method (details are provided in the Methods). A common feature to both structures is the large distance ($\sim100-200\;\mu$m) existing between the gate electrode and the ISFET channel, which prevents any conventional electron leakage current through the substrate (sapphire or silicon dioxide) stemming from the application of the polarization voltage to the electrolyte ($|V_g|\leq5.5$ V for DEME-TFSI). 
Type-A and Type-B nanotransistors show the typical hysteretic current-voltage ($IV$) characteristic of diffusive superconducting weak-links \cite{Courtois2008} (Fig. \ref{Fig1}, c and d), where the transition current from the normal to the superconducting state is known as retrapping current ($I_R$). As expected, $I_C$ and $I_R$ decrease monotonically with bath temperature ($T$) until their full suppression at $T_C$ (Fig. \ref{Fig1}, e and f). In particular, the devices show $T_{C,A}\sim7.3$ K and $T_{C,B}\sim7.6$ K for Type-A and Type-B structure, respectively. In addition, $I_C$ follows the Bardeen theoretical curve \cite{Bardeen1962} (dashed lines), where $I_{C_A,0}\sim500\;\mu$A and $I_{C_B,0}\sim460\;\mu$A are the zero-temperature critical currents obtained from the fit of Type-A and Type-B devices, respectively.

\begin{figure*} [t!]
        \begin{center}
                \includegraphics{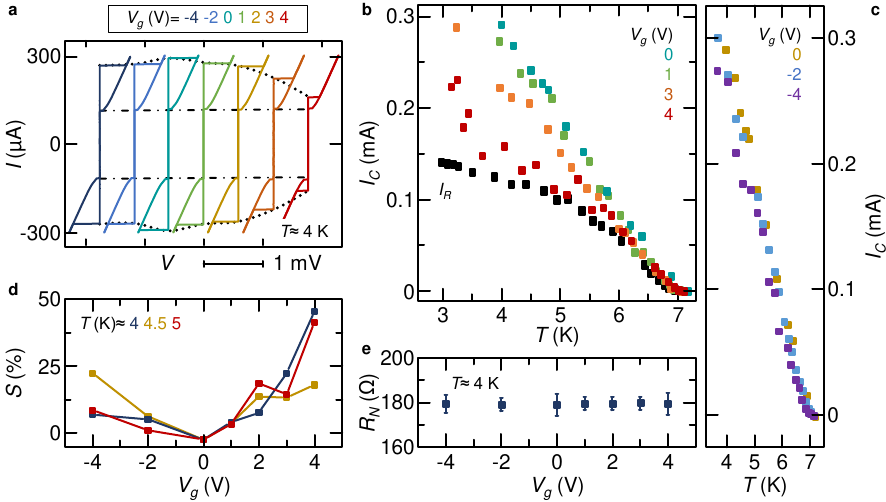}
        \end{center}
        \caption{\label{Fig2}\textbf{Transport properties of a typical Type-A Nb ISFET.} 
        (\textbf{a}) $IV$ characteristics measured for different values of gate voltage at $T\sim4$ K. The curves are horizontally offset for clarity. Dotted black lines are guides for the eye highlighting the $I_C$ evolution under $V_g$, whereas dash-dotted lines mark the independence of $I_R$ from gating. 
        (\textbf{b}) Temperature dependence of $I_C$ measured for different positive values of $V_g$. A typical $I_R$ vs $T$ characteristics is also shown. 
         (\textbf{c}) Temperature dependence of $I_C$ measured for different negative values of $V_g$.
        (\textbf{d}) Supercurrent suppression parameter $S$ versus $V_g$ for selected values of $T$.
        (\textbf{e}) Constriction normal-state resistance $R_N$ versus $V_g$ at $T\sim4$ K.
        }
\end{figure*}

We now focus on the electrostatic control of the supercurrent in a  Type-A ISFET consisting of a Josephson nanoconstriction (Fig. \ref{Fig1}b). To this end, we measured the $IV$ characteristics as a function of bath temperature for different values of $V_g$. Since the ions in the electrolyte are frozen and immobile at cryogenic temperatures, we applied $V_g$ at room temperature before cooling down the devices (details are provided in the Supplementary Information). 
Figure \ref{Fig2}a shows the $IV$ characteristics recorded at $T\simeq4$ K for different values of $V_g$. The critical current decreases with the amplitude of the gate voltage for both positive and negative polarities of $V_g$ (dotted lines), while $I_R$ turns out to be unaffected by gating effect (dash-dotted lines). We emphasize that at $V_g=4$ V we also disconnected the gate voltage source at cryogenic temperatures (i.e., when the ions are frozen in a fixed configuration) yielding no variation on the suppression of $I_C$, thereby further excluding any possible relation between the critical current suppression and the leakage currents. Despite the characteristic penetration depth of the electric field into a metal ranges from 2 \AA\; to 4 nm \cite{Piatti2017}, any perturbation of the superconducting condensate propagates into the bulk for a distance a several times the coherence length ($\xi$) \cite{DeSimoni2018}. Since in Nb thin films $\xi$ can reach up to 30 nm \cite{Draskovic2013}, we can assume that the effect of an electric field on the ISFET surface can influence the complete thickness of the device.

The general behavior of $I_C$ as a function of $T$ is shown in Fig. \ref{Fig2}b for selected positive values of $V_g$. The critical current is strongly suppressed by $V_g$ at low temperature, while all the curves collapse into the zero-gate trace for $T\to T_C$. 
This behavior resembles the temperature dependence of the critical current measured in conventionally-gated Ti nano constrictions \cite{Paolucci2019}.
Furthermore, in full agreement with previous experiments on Nb Josephson nano-bridges \cite{Puglia2020_2}, this unconventional gating has no effect on the critical temperature of the nanosized superconductor. We stress that $I_C$ is also suppressed for negative values of the gate voltage (Fig. \ref{Fig2}c). The bipolarity of gating effect can be quantified via the \textit{suppression} parameter, defined as $S(T,V_g)=100\times[I_C(T,0)-I_C(T,V_g)]/I_C(T,0)$. 
The $S$ parameter increases with the magnitude of $V_g$ for both polarities, reaching its maximum of $\sim 45\%$ at $V_g=4$ V and $T\simeq4$ K (Fig. \ref{Fig2}d). Remarkably, the normal-state resistance of Type-A ISFETs is unaffected by the gating (Fig. \ref{Fig2}e). These features suggest that changes of the charge concentration are not driving the supercurrent suppression in our devices. Therefore, this excludes conventional charge accumulation/depletion due to field-effect or electrochemical modification of the device surface. In addition, the electrolyte/surface chemical interaction results in evident signatures in the ionic current ($I_g$) flowing during the electrolyte polarization at room temperature \cite{Kirby2013}. Our chornocoulometry measurements show the typical traces of the EDL creation in the absence of chemical interaction (see Supplementary Information). Therefore, we can exclude electrochemical surface modifications as the origin of the effect.

\begin{figure*} [t!]
        \begin{center}
                \includegraphics{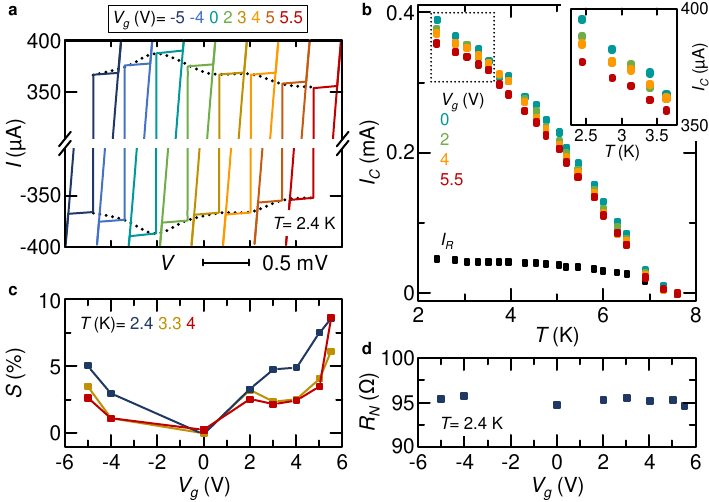}
        \end{center}
        \caption{\label{Fig3}\textbf{Transport properties of a typical Type-B Nb ISFET.} 
        (\textbf{a}) $IV$ characteristics measured for different values of gate voltage at $T\sim4$ K. The curves are horizontally offset for clarity. The dotted black lines are guides for the eye highlighting $I_C$ evolution with gating.
        (\textbf{b}) Temperature dependence of $I_C$ measured for different positive values of $V_g$. A typical $I_R$ vs $T$ curve is also shown. (Inset) Blow up of main panel for $T<3.750$ K.
        (\textbf{c}) Suppression parameter $S$ versus $V_g$ for selected values of $T$.
        (\textbf{d}) Nanowire normal-state resistance $R_N$ versus $V_g$ at $T=2.4$ K.
        }
\end{figure*}

To examine whether this unconventional gating effect is general or related to a specific sample geometry, we performed similar experiments on Type-B ISFETs, where the superconducting channel consists of a long Nb nanowire (Fig. \ref{Fig1}b). Notably, the overall qualitative behavior  of this structure is analogous to Type-A sample. Indeed, the $IV$ characteristics of a typical Type-B ISFET recorded at $T=2.4$ K show bipolar $I_C$ suppression with the gate voltage (Fig. \ref{Fig3}a). At a given value of $V_g$, the supercurrent is strongly reduced at low temperatures, whereas it approaches the zero-gate value by increasing the temperature (Fig. \ref{Fig3}b). Furthermore, the superconducting critical temperature and the retrapping current are $V_g$-independent. As a consequence of the $I_C(V_g)$ versus $T$ characteristics, the suppression parameter decreases with temperature for both polarities of gate voltage (Fig. \ref{Fig3}c). 
Similarly to Type-A ISFET (Fig. \ref{Fig2}d),
gating effect is more pronounced for positive values of gate voltage ($S\sim9\%$ at $V_g=5.5$ V), due to the higher polarization efficiency of the employed electrolyte (DEME-TFSI) \cite{Sato2004}. 
Finally, the normal-state resistance of the nanotransistor channel is not affected by gating, since $R_N$ does not show any clear trend with $V_g$ (Fig. \ref{Fig3}d). 

The phenomenology shown by Type-A and Type-B ISFETs fully resembles the effect of conventional solid gating observed so far on nanosized superconducting systems \cite{DeSimoni2018,Paolucci2018,DeSimoni2019,Rocci2020,Elalaily2021,Paolucci2019_2,Puglia2020,Paolucci2019,Puglia2020_2}.
The absence of full supercurrent suppression in the ISFETs might be ascribed to the profile of the electric field generated by ionic gating. Indeed, lateral solid gates tend to produce localized and strongly anisotropic electric fields \cite{DeSimoni2019,Rocci2020}, while electrolytes tend to generate rather isotropic electric field profiles on the nanotransistors channel surface.
Since ionic gating 
avoids any leakage current, we conclude that the observed supercurrent suppression is exclusively driven by the electrostatic field therefore ruling out any quasiparticle injection-related mechanism at the origin of the effect \cite{Ritter2021,Catto2021,Golokolenov2021,Alegria2021,Ritter2021_2}. 
Moreover, the observed bipolarity  together with the invariance of $R_N$ and $T_C$ also exclude significant modulations of the superconductor chemical potential due to charge accumulation/depletion (due to conventional field-effect or electrochemistry) which, differently, were reported in ionic-gated micro-scale metals in the normal \cite{Daghero2012} and superconducting \cite{Choi2014,Piatti2021} state. 
This difference might stem from the nanoscale channel of our ISFETs as compared to the micro-scale one of other works on Nb thin films, or from the superconducting film thickness \cite{Choi2014}.
As a consequence of the chemical potential steadiness, this unconventional field-effect seems not to act directly on the amplitude of the pairing potential but suggests a non-trivial interaction with the superconducting macroscopic phase \cite{Paolucci2019_2,Puglia2020}. 
Since the microscopic nature of such interaction is still almost entirely unknown, further experimental and theoretical investigations are required to shed light on this gating effect. 

\begin{figure} [t!]
        \begin{center}
                \includegraphics{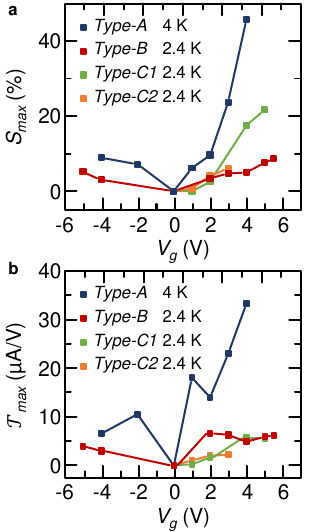}
        \end{center}
        \caption{\label{Fig4}\textbf{Performance of different Nb ISFETs.} 
        (\textbf{a}) Maximum value of the supercurrent suppression parameter versus $V_g$ for four different devices.
        (\textbf{b}) Maximum value of the transconductance versus $V_g$ for the same devices shown in panel a. In both panels, 
        each device refers to a single temperature, as indicated in the legend.
        }
\end{figure}

We wish to finally comment the performance of our Nb ISFETs from the application point of view. 
To this end, it is interesting to compare the maximum suppression parameter obtained in various ISFETs. 
Figure \ref{Fig4}a shows $S_{max}$ for different ISFET geometries. In particular, samples discussed so far are also compared with an additional set of Type-C devices, that are Josephson nano-constrictions realized by a bottom-up approach on top of a silicon dioxide substrate (details are provided in Methods). The suppression parameter seems to be generally larger for devices fabricated with a bottom-up approach (Type-A and Type-C), since possible tiny resist residues originating from the etching mask might increase the distance between the charge planes forming the EDL at the surface of the superconducting channel, thereby yielding a reduced electric field.
Moreover, in light of possible applications, it is worth to discuss the \textit{transconductance} of the ISFETs, a  key figure of merit for field-effect transistors. For the ISFETs, it can be defined as $\mathcal{T}=\delta I_C/|V_g|$, where $\delta I_C=I_C(V_g)-I_C(0)$. 
Figure \ref{Fig4}b shows the maximum value of $\mathcal{T}$ versus $V_g$ deduced for four different ISFETs. Notably, for all our ionic-gated devices, $\mathcal{T}_{max}$ (which obtains values as high as $\sim 35\;\mu$A/V) is comparable to the transconductance achieved so far in conventional superconducting \cite{Nishino1989}, Josephson \cite{Clark1980}, and all-metallic field-effect nanotransistors \cite{Paolucci2018}. 

In conclusion, our work provides the clear and unambiguous proof of the non-conventional and disruptive impact of intense electrostatic fields on the supercurrent flow in nano-sized metallic superconductors.
The use of ionic gating rules out both quasiparticle overheating and any other charge injection-based mechanism at the origin of the effect, since electron transfer across the EDL is prohibited at cryogenic temperatures.
Moreover, the bipolar supercurrent suppression observed in ISFETs of different geometries, along with the absence of critical temperature and normal-state resistance variations, exclude conventional charge accumulation/depletion due to field-effect or electrochemical modification of the Nb surface. Therefore, our results advise for further theoretical and experimental investigations to unveil the microscopic genesis of field-effect in metallic superconductors. From the technological side, our ISFETs might be at the basis of innovative combined ionic-solid-gating devices suitable for a number of different  applications in leading-edge superconducting electronics, quantum computing, and sensing.

\section*{Methods}
\subsection*{Fabrication protocols}
\textbf{Type-A ISFET}s were fabricated by bottom-up approach on a sapphire substrate. A single-step electron-beam lithography (EBL) was realized to pattern a polymethyl-methacrylate (PMMA, thickness $\sim250$ nm) evaporation mask. To avoid charging of the insulating substrate, a conductive polymer (Elettra 92) was spin-coated on the PMMA layer before the EBL.
After the development of the resist mask, a 5-nm-thick Ti adhesion layer was dc-sputtered before the superconducting Nb film of thickness $t_A\sim50$ nm. After the lift-off procedure, the resulting nanotransitor channel is a nanoconstriction Josephson junction of length $l_A\sim300$ nm and width $w_A\sim100$ nm. 
The gate counter electrode is distant $\sim100\;\mu$m from the transistor channel.
After wire-bonding the device on a suited chip carrier, a drop of diethylmethyl(2-methoxyethyl)ammonium bis(trifluoromethylsulfonyl)imide (DEME-TFSI) was placed on the device.

\textbf{Type-B ISFET}s were fabricated by top-down approach on a sapphire substrate. A 5-nm-thick Ti adhesion layer was dc-sputtered before the superconducting Nb film of thickness $t_B\sim50$ nm. A single-step EBL was realized to pattern the PMMA (thickness $\sim250$ nm) etching mask. 
After the development, a step of inductively coupled plasma reactive ion etching [ICP-RIE, BCl$_3$ (8 sccm)/Cl$_2$ (6 sccm)/Ar (10 sccm)] etches the Nb film defining the device.
The resulting nanotransitor channel is a nanowire of length $l_B\sim1\;\mu$m and width $w_B\sim100$ nm. 
The gate counter electrode is distant $\sim100\;\mu$m from the transistor channel.
After wire-bonding the device on a suited chip carrier, a drop of DEME-TFSI was placed on the device.

\textbf{Type-C ISFET}s were fabricated by bottom-up approach on a silicon substrate covered by 300 nm of thermally grown dry silicon dioxide. A single-step EBL was realized to pattern a polymethyl-methacrylate (PMMA, thickness $\sim250$ nm) evaporation mask.
After the development of the resist mask, a 5-nm-thick Ti adhesion layer was dc-sputtered before the superconducting Nb film of thickness $t_C\sim50$ nm. The resulting nanotransitor channel is a nanoconstriction Josephson junction of length $l_C\sim300$ nm and width $w_C\sim125$ nm. 
The gate counter electrode is distant $\sim100\;\mu$m from the transistor channel.
After wire-bonding the device on a suited chip carrier, a drop of DEME-TFSI was placed on the device.

\subsection*{Measurement set-up}
The measurements were performed from room temperature down to 2.4 K in a dry cryostat (OptistatDry, Oxford Instruments) equipped with RC-fiters of resistance $\sim1.15$ k$\Omega$. The gate voltage was applied by means of a semiconductor characterization system (Keithley 4200). 
The current versus voltage characteristics of the ISFETs were obtained by applying a biasing current through a low noise source (GS200, Yokogawa) and measuring the voltage drop by a room-temperature battery-powered differential preamplifier (Model 1211, DL Instruments). 
Alternatively, the device characterization was performed by a 750-KHz bandwidth input/output analog-to-digital/digital-to-analog converter (ADC/DAC, National Instruments) board for the generation of the bias current and the acquisition of the voltage drop signal arising from a room-temperature battery-powered differential preamplifier (Model 1211, DL Instruments). 
In both cases, we performed several repetitions of the IV characteristics (up to 50 depending on the device).

\section*{Data availability}
All other data that support the plots within this paper and other findings of this study are available from the corresponding author upon reasonable request.

\section*{Acknowledgements}
The authors acknowledge the European Research Council under Grant Agreement No. 899315 (TERASEC), and the EU’s Horizon 2020 research and innovation program under Grant Agreement No.  800923 (SUPERTED) and No. 964398 (SUPERGATE) for partial financial support.

\section*{Author contributions} \label{sec:Author contributions}
F.P., G.D.S., L.B. and C.P. fabricated the samples. F.P., F.C. and L.B. performed the experiments with inputs from  G.D.S., E.S, S.R. and F.G.. F.P. and F.C. analysed the experimental data with inputs from F.G.. F.P. and F.G. wrote the manuscript with inputs from all authors. F.G. supervised and coordinated the project. All authors discussed the results and their implications equally at all stages.

\section*{Additional information}
The authors declare no competing financial interests.

\section*{Supporting Information}
This material is available free of charge via the internet at http://pubs.acs.org.

\newpage

\section*{Supplementary Information for Electrostatic field-driven supercurrent suppression in ionic-gated metallic superconducting nanotransistors}
\section*{Polarization of the electrolyte}
\begin{figure*} [t!]
        \begin{center}
                \includegraphics{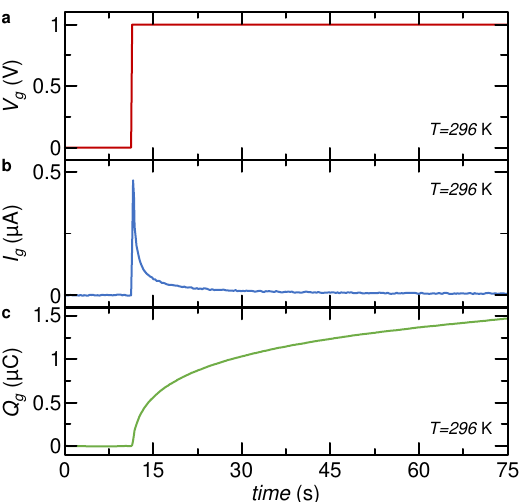}
        \end{center}
        \caption{\label{FigS1}\textbf{Typical charging curve of the electrolyte.} 
        (\textbf{a}) Voltage applied to the gate counter electrode as a function of time.
        (\textbf{b}) Ionic current flowing into the electrolyte as a function of time due to the $V_g$ profile shown in panel a.
        (\textbf{c}) Charge displaced in the electrolyte as a function of time corresponding to the ionic current shown in panel b.
        }
\end{figure*}

Figure \ref{FigS1} shows a typical charging characteristic of our IJoFETS recorded at room temperature ($T=296$ K) before cooling down the devices to cryogenic temperatures for the electronic transport experiments. When a gate voltage is applied (panel a, $V_g=1$ V), a peak in the ionic current appears (panel b, $I_g$) due to the formation of the electric double layers (EDLs) at the interfaces of the electrolyte with the nanotransistor channel and the gate counterelectrode. On the one hand, $I_g$ shows a fast decay with time. On the other hand, an ionic current several orders of magnitude lower than the peak value is always present, since a charge flow is necessary into the electrolyte to maintain the gradient of ionic concentration \cite{Yuan2010}. We note that thus $I_g$ vs time characteristics is typical of the formation of an EDL without any chemical electrolyte/surface interaction \cite{Kirby2013}.

The ionic charge displaced in the electrolyte is given by
\begin{equation}
    Q_g(t)=\int\limits_0^t I_g(t') \mathrm{d}t',
\end{equation}
where $t$ is the time. Figure \ref{FigS1}c shows the typical time dependence of $Q_g$: an exponential charging with time during the EDLs formation and, later, a square dependence with time ($t^2$) due to electrochemical effects in the electrolyte bulk \cite{Kirby2013}.

\section*{Electronic transport in Type-C IJ\lowercase{o}FETs}
\begin{figure*} [t!]
        \begin{center}
                \includegraphics{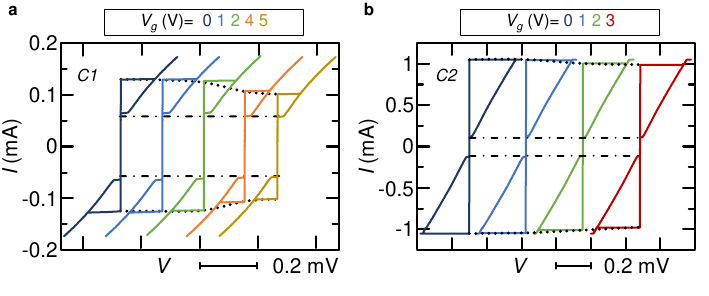}
        \end{center}
        \caption{\label{FigS2}\textbf{IV characteristics for Type-C IJoFETs.} 
        $IV$ characteristics measured for different values of gate voltage at $T=2.4$ K for devices C1 (panel \textbf{a}) and C2 (panel \textbf{b}). The curves are horizontally offset for clarity. Dotted black lines are guides for the eye highlighting the $I_C$ evolution under $V_g$, whereas dash-dotted lines mark the independence of $I_R$ from gating. 
        }
\end{figure*}

We present the electrostatic control of the supercurrent in Type-C IJoFETs consisting of a Josephson nanoconstriction fabricated with top-down approach on a silicon substrate covered with 300 nm of thermally grown silicon dioxide.
Since the electrolyte can be polarized only at temperatures higher than the electrolyte glass transition temperature, we applied the gate voltage at room temperature and, then, we cooled down the devices to cryogenic temperatures to investigate the behavior of the supercurrent with temperature while a strong electrostatic field is present. To this end, we recorded the $IV$ characteristics at different values of $V_g$ as a function of temperature.
Figure \ref{FigS2} shows the $IV$ characteristics recorded at $T=2.4$ K for different values of gate voltage of devices C1 (panel a) and C2 (panel b), respectively. The critical current decreases with the amplitude of the gate voltage (dotted black lines), while the retrapping current ($I_R$) \cite{Courtois2008} turns out to be almost unaffected by gating effect (dashed-dotted black lines). In particular, $I_C$ decreases of $\sim6\%$ at $V_g=5$ V for sample C1, while it is suppressed of about $3\%$ at $V_g=3$ V for sample C2.

\section*{Devices parameters}
Based on the device structure and transport properties, we can estimate several parameters of the superconducting Nb film, such as the normal-stare resistance ($R_N$), the critical temperature ($T_C$), the zero-temperature Bardeen-Cooper-Schrieffer (BCS) energy gap ($\Delta_0$), and the zero-temperature critical current ($I_{C,0}$).

The normal-state resistance is obtained by a linear fit of the current versus voltage characteristics in the dissipative state. The error bars arise from the averages between the values of $R_N$ obtained from the fit procedure on the different repetitions.

The zero-temperature BCS superconducting energy gap can be calculated from
\begin{equation}
    \Delta_0=1.764\, k_B\, T_C,
\end{equation}
where $k_B$ is the Boltzmann constant and $T_C$ is the superconducting critical temperature. 



The phenomenological Bardeen model for the temperature dependence of the superconducting critical current reads \cite{Bardeen1962}

\begin{equation}
  I_C(T)=I_{C,0}\bigg[ 1-\bigg(\frac{T}{T_C}\bigg)^2\bigg]^{3/2},
\end{equation}
where $I_{C,0}$ is the zero-temperature critical current.\\

The basic parameters deduced for every device are resumed in the following table.

\begin{center}
\begin{tabular}{|p{2cm}|p{2cm}|p{2cm}|p{2cm}|}
 \hline
 \multicolumn{4}{|c|}{\textbf{DEVICES PARAMETERS}} \\
 \hline
 Sample   &   $T_C$ [K]   &    $R_N$ [$\Omega$]    & $\Delta_0$ [$\mu$eV] \\
 \hline
 \textbf{Type-A}    &   7.3 &   180     &   1.11    \\
 \textbf{Type-B}    &   7.6 &   94.6    &   1.15    \\
 \textbf{Type-C1}   &   5.4 &   18      &   0.82    \\
 \textbf{Type-C2}   &   6.4 &   28.8    &   0.97    \\
 \hline
\end{tabular}
\end{center}

\end{document}